\documentclass[lettersize,journal]{IEEEtran}
\usepackage{amsmath,amssymb,graphicx}
\usepackage{booktabs}
\usepackage{lmodern}
\usepackage[T1]{fontenc}
\usepackage{microtype}
\usepackage{comment}
\usepackage{xcolor}
\usepackage{cite}
\usepackage{algorithm,algpseudocode}
\usepackage{array}
\usepackage{textcomp}
\usepackage{url}
\usepackage{hyperref}
\usepackage{multirow}
\usepackage{caption}
\usepackage[flushleft]{threeparttable}
\usepackage{epstopdf}
\usepackage{subcaption}
\usepackage[export]{adjustbox}
\usepackage{wrapfig}

\hypersetup{
    colorlinks=true,
    citecolor=black,
    linkcolor=blue,
    filecolor=magenta,      
    urlcolor=cyan,
}

\newcommand{\orcidauthorLuqman}{\href{https://orcid.org/0000-0001-5792-0842}{\includegraphics[scale=0.05]{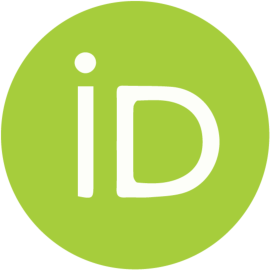}}}
\newcommand{\orcidauthorFazal}{\href{https://orcid.org/0000-0003-0405-0083}{\includegraphics[scale=0.05]{orcid_16x16.eps}}}
\newcommand{\orcidauthorSaif}{\href{https://orcid.org/0000-0002-7262-183X}{\includegraphics[scale=0.05]{orcid_16x16.eps}}}
\newcommand{\orcidauthorIrfan}{\href{https://orcid.org/0000-0003-4161-6875}{\includegraphics[scale=0.05]{orcid_16x16.eps}}}

\newcommand{\orcidauthorMaaz}{\href{https://orcid.org/0009-0000-2971-3352}{\includegraphics[scale=0.05]{orcid_16x16.eps}}}

\newcommand{\orcidauthorSaid}{\href{https://orcid.org/0009-0007-8491-4959}{\includegraphics[scale=0.05]{orcid_16x16.eps}}}

\newcommand{\orcidauthorJalal}{\href{https://orcid.org/0000-0001-6092-705X}{\includegraphics[scale=0.05]{orcid_16x16.eps}}}

\newcommand{\orcidauthorSalman}{\href{https://orcid.org/0009-0007-1602-2641}{\includegraphics[scale=0.05]{orcid_16x16.eps}}}

\begin{document}

\title{Distributed Deep Learning with RIS Grouping for Accurate Cascaded Channel Estimation}
\author{%
\raggedright
Saifur~Rahman$^{2,a}\orcidauthorSaif{}$,
Syed~Luqman~Shah$^{3,b}\orcidauthorLuqman{}$,
Salman~Khan$^{1,c}\orcidauthorSalman{}$,
Jalal~Khan$^{4,d}\orcidauthorJalal{}$,
Muhammad~Irfan$^{2,e}\orcidauthorIrfan{}$,
Maaz~Shafi$^{1,f}\orcidauthorMaaz{}$,
Said~Muhammad$^{1,g}\orcidauthorSaid{}$,
Fazal~Muhammad$^{1,h}\orcidauthorFazal{}$,
Mohammad~Shahed~Akond$^{5,*}$
\\[1.75em]
\textbf{AFFILIATIONS}\\[1em]
$^{1}$Department of Electrical Engineering, University of Engineering and Technology (UET) Mardan, 23200, Pakistan\\
$^{2}$Electrical Engineering Department, College of Engineering, Najran University, Najran 61441, Saudi Arabia\\
$^{3}$Centre for Wireless Communications (CWC), University of Oulu, 90570 Oulu, Finland\\
$^{4}$Department of Telecommunication Engineering, University of Engineering and Technology (UET) Mardan, 23200, Pakistan\\
$^{5}$Department of Electrical and Electronic Engineering, Khwaja Yunus Ali University, Bangladesh\\[1.75em]
$^{a}$\texttt{srrahman@nu.edu.sa}\\
$^{b}$\texttt{syed.luqman@oulu.fi}\\
$^{c}$\texttt{engineersalmank@gmail.com}\\
$^{d}$\texttt{jalal@uetmardan.edu.pk}\\
$^{e}$\texttt{irfan16.uetian@hotmail.com}\\
$^{f}$\texttt{i.maazsk@gmail.com}\\
$^{g}$\texttt{engineersaidm@gmail.com}\\
$^{h}$\texttt{fazal.muhammad@uetmardan.edu.pk}\\
$^{*}$\textbf{Author to whom correspondence should be addressed:} \texttt{msakond.eee@kyau.edu.bd}\\[0.25em]
\par}

\maketitle

\begin{abstract}
Reconfigurable Intelligent Surface (RIS) panels are envisioned as a key technology for sixth-generation (6G) wireless networks, providing a cost-effective means to enhance coverage and spectral efficiency. A critical challenge is the estimation of the cascaded base station (BS)-RIS-user channel, since the passive nature of RIS elements prevents direct channel acquisition, incurring prohibitive pilot overhead, computational complexity, and energy consumption. To address this, we propose a deep learning (DL)-based channel estimation framework that reduces pilot overhead by grouping RIS elements and reconstructing the cascaded channel from partial pilot observations. Furthermore, conventional DL models trained under single-user settings suffer from poor generalization across new user locations and propagation scenarios. We develop a distributed machine learning (DML) strategy in which the BS and users collaboratively train a shared neural network using diverse channel datasets collected across the network, thereby achieving robust generalization. Building on this foundation, we design a hierarchical DML neural architecture that first classifies propagation conditions and then employs scenario-specific feature extraction to further improve estimation accuracy. Simulation results confirm that the proposed framework substantially reduces pilot overhead and complexity while outperforming conventional methods and single-user models in channel estimation accuracy. These results demonstrate the practicality and effectiveness of the proposed approach for 6G RIS-assisted systems.
\end{abstract}


\section{Introduction}
Reconfigurable intelligent surface (RIS) panels have emerged as a transformative technology for beyond-fifth-generation (B5G) and sixth-generation (6G) wireless systems.
These panels are low-cost, nearly passive metasurfaces that can programmatically control the wireless propagation environment~\cite{b1,b4}.
RIS is especially beneficial in millimeter-wave (mmWave) and terahertz (THz) bands, where blockage and severe path loss are common.
By intelligently redirecting reflections, RIS can extend coverage and improve spectral efficiency without requiring active radio-frequency (RF) chains on the surface~\cite{b2,b3,b5}. RIS can also create a line-of-sight (LoS) link for users that are otherwise in non-LoS conditions.
Such LoS links can further enhance localization accuracy by providing stronger and more stable reference signals~\cite{ismailc, ismailj}. They also enable higher received signal strength and lower path loss, which improve link reliability and coverage at the cell edge~\cite{d2dPaper}. In addition, robust LoS connectivity supports more flexible and dense deployments of RIS panels and access points in next-generation wireless systems. However, all these benefits critically rely on the acquisition of accurate channel state information (CSI)~\cite{b19}.

A primary challenge is to obtain CSI for the cascaded link that involves the base station (BS), the RIS, and the user. Because RIS elements are passive, the BS–RIS and RIS–user channels cannot be estimated independently.
Instead, only their cascaded product is observable at the receiver. In frequency-division duplexing (FDD) systems, where uplink-downlink reciprocity does not hold, downlink pilots must activate the RIS, and the user measures the combined reflected signals. For an RIS with $N$ elements and a BS with $M$ antennas, the cascaded channel comprises $NM$ complex parameters. Traditional least-squares (LS) estimation demands at least $Q \geq NM$ pilot symbols (or $Q \geq N'M$ with grouping, where $N' < N$), which becomes impractical for large $N$ due to excessive overhead~\cite{b6,b7,b8,b9,b10}.
Advanced estimators exploit channel structures, such as angular sparsity and spatial correlations, to reduce pilot requirements~\cite{b19}. Deep neural networks can map low-dimensional pilot observations to the full high-dimensional cascaded channel, often integrating compressive sensing or denoising techniques~\cite{b11,b12}. Nevertheless, models trained in isolated user or scenario-specific settings frequently exhibit poor generalization to varying user locations, blockage conditions, or angular spreads. This is exacerbated by non-independent and identically distributed (non-IID) data across a cell, leading to degraded accuracy and reliability~\cite{b11,b12,b14}.
Distributed machine learning (DML), including federated learning (FL), tackles non-IID data by aggregating model updates from distributed devices without centralizing raw data, as in the Federated Averaging (FedAvg) algorithm~\cite{b17}. For heterogeneous environments, specialization enhances performance: sparsely-gated mixture-of-experts (MoE) models use a learned gate to route inputs to a subset of experts, balancing accuracy and efficiency~\cite{b18}. In RIS CSI estimation, a gate based on coarse scenario indicators (e.g., elevation sectors) can select region-specific experts, avoiding generic feature extraction. Recent developments, such as transformer-based CSI prediction~\cite{b15} and comprehensive surveys~\cite{b14}, emphasize the need for estimators that (i) function with few pilots ($Q$), (ii) generalize across regions, and (iii) maintain low online computational cost at user equipment (UE).
In this paper, we present a DML framework for downlink cascaded channel estimation under limited pilot regimes. Our approach integrates three key components: (i) RIS grouping to compress the effective cascaded channel from $NM$ to $N'M$ parameters (with $N' = N/g$ for group size $g$), thereby reducing pilot and computational demands; (ii) a region-gated MoE architecture that classifies the propagation region first and then activates only the relevant expert for feature extraction from the pilot data; and (iii) synchronous FedAvg involving the BS and users to train a shared estimator on diverse, non-IID data without exchanging raw samples~\cite{b17}. The gate employs sparse routing inspired by~\cite{b18}, ensuring that only one expert is active during inference, maintaining UE runtime comparable to a single compact convolutional neural network (CNN).

\subsection{Related Work}
\label{subsec:related}
Initial studies on RIS-assisted systems adapted LS and minimum mean-squared error (MMSE) estimators to the cascaded BS-RIS-user channel in FDD setups, incorporating custom pilot sequences or structural priors~\cite{b6,b8,b9,b10}. Although effective for modest scales, these methods necessitate $Q \geq NM$ pilots (or $Q \geq N'M$ with grouping) for reliable LS solutions and require precise covariance knowledge for MMSE, which is challenging in large-scale or dynamic settings. Our framework specifically targets this pilot overhead bottleneck through short-pilot operations.

Deep learning-based estimators minimize pilot needs by learning nonlinear mappings from reduced-dimensional pilots to the full cascaded channel. Examples include ordinary differential equation (ODE)-inspired CNNs that exploit spatial correlations for channel extrapolation~\cite{b11} and denoising networks paired with compressive sensing for noise suppression in mmWave bands~\cite{b12}. While these yield significant data-driven improvements, they are often trained in single-scenario contexts and suffer performance drops under distribution shifts (e.g., varying angular spreads or blockages). We counter this with region-aware specialization and distributed training.

To handle heterogeneity across users and regions without aggregating raw data, DML aggregates model parameters from devices using algorithms like FedAvg~\cite{b17, b20}. In RIS contexts,~\cite{b13} proposed a DML scheme for downlink channel estimation, demonstrating enhanced generalization over per-user training. Compared to~\cite{b13}, our work adds (i) RIS grouping to shrink the parameter space from $NM$ to $N'M$, cutting pilot and compute costs; and (ii) a region-gated design that engages only one expert per inference, reducing on-device complexity while preserving accuracy.

Sparsely-gated MoE models direct inputs to select experts via a trainable gate, optimizing capacity and efficiency~\cite{b18}. We adapt this to the physical layer with a lightweight scenario classifier as the gate and $R$ region-specific experts for extraction; only the chosen expert runs at the UE, keeping latency akin to a single small CNN (see Sec.~\ref{sec:proposed-dml}).

Transformer-based methods for CSI prediction and deep-unfolding for RIS phase optimization have gained traction~\cite{b15}, with surveys outlining ML pipelines for RIS CSI acquisition~\cite{b14}. These advancements underscore the demand for estimators that (i) perform well with small $Q$, (ii) adapt to non-IID spatial variations, and (iii) minimize UE inference overhead—core objectives of our design.
Relative to classical LS/MMSE~\cite{b6,b8,b19} and standalone deep estimators~\cite{b11,b12}, our framework merges DML (FedAvg)~\cite{b17} with gated specialization~\cite{b18} and RIS grouping to concurrently lower pilot overhead, boost cross-region robustness, and control online complexity. We formalize the observation models, identifiability, and grouping in Sec.~\ref{sec:system-model}, and detail the gated DML architecture and protocols in Sec.~\ref{sec:proposed-dml}.

\subsection{Contributions} 
Our key contributions, building on existing literature, include:
\begin{itemize}
\item A deep learning estimator for cascaded CSI that reconstructs the channel from far fewer pilots ($Q \ll NM$ or $Q \ll N'M$ with grouping), using a linear observation model aligned with RIS physics and unit-modulus phases~\cite{b6,b8}.
\item Formalization of RIS grouping via a selection/aggregation operator that reduces $N$ elements to $N'$ control units, maintaining energy conservation and enabling lower-dimensional estimation.
\item A region-gated MoE architecture where a lightweight classifier routes inputs to a specific regional expert; only one expert activates online, providing specialization without increasing latency~\cite{b18}.
\item {\color{black}A synchronous FedAvg-based DML protocol involving users and the BS to train the shared model and experts on local non-IID data, enhancing generalization across regions~\cite{b17,b13}.}
\item Detailed complexity analysis with closed-form multiply-accumulate (MAC) counts for the active expert and mapping head, plus empirical evidence showing superior NMSE over LS/MMSE at reduced $Q$.
\end{itemize}

\subsection{Scope and organization} 
This work concentrates on estimating the cascaded BS-RIS-user channel in FDD systems; the direct BS-user link, if available, can be estimated separately by disabling the RIS~\cite{b3}. Sec.~\ref{sec:system-model} presents the system and observation models, including grouping and identifiability conditions. Sec.~\ref{sec:proposed-dml} describes the proposed DML estimator, training, and inference procedures. Sec.~\ref{sec:simulation} outlines simulation setups and comparisons with LS and MMSE baselines, followed by conclusions in Sec.~\ref{sec:conclusion}.

\begin{table}[htbp]
\caption{Mathematical Notation Used in This Paper}
\label{tab:notation}
\centering
\begin{tabular}{@{}ll@{}}
\toprule
\textbf{Notation} & \textbf{Description} \\
\midrule
$\mathbf{a}$, $\mathbf{b}$ & Bold lowercase letters denote column vectors \\
$\mathbf{A}$, $\mathbf{B}$ & Bold uppercase letters denote matrices \\
$\mathbf{A}^{\mathrm{T}}$ & Transpose of $\mathbf{A}$ \\
$\mathbf{A}^{\mathrm{H}}$ & Hermitian (conjugate transpose) of $\mathbf{A}$ \\
$\operatorname{diag}(\mathbf{a})$ & Diagonal matrix with entries of $\mathbf{a}$ on its diagonal \\
$\|\mathbf{a}\|_2$ & Euclidean ($\ell_2$) norm of vector $\mathbf{a}$ \\
$\mathbf{A} \otimes \mathbf{B}$ & Kronecker product of matrices $\mathbf{A}$ and $\mathbf{B}$ \\
$\mathrm{vec}(\mathbf{A})$ & Vectorization of matrix $\mathbf{A}$ by stacking its columns \\
$\sigma_n^2$ & Noise variance of AWGN samples \\
$\mathrm{SNR}$ & Signal-to-noise ratio (SNR), defined as $1/\sigma_n^2$ 
\\
$\mathcal{O}(\cdot)$ & Big-$\mathcal{O}$ notation for computational complexity \\
\bottomrule
\end{tabular}
\end{table}

\vspace{1em}

\section{System Model and Problem Formulation}
\label{sec:system-model}

\begin{figure}[tbp]
\centering
\includegraphics[width=\linewidth]{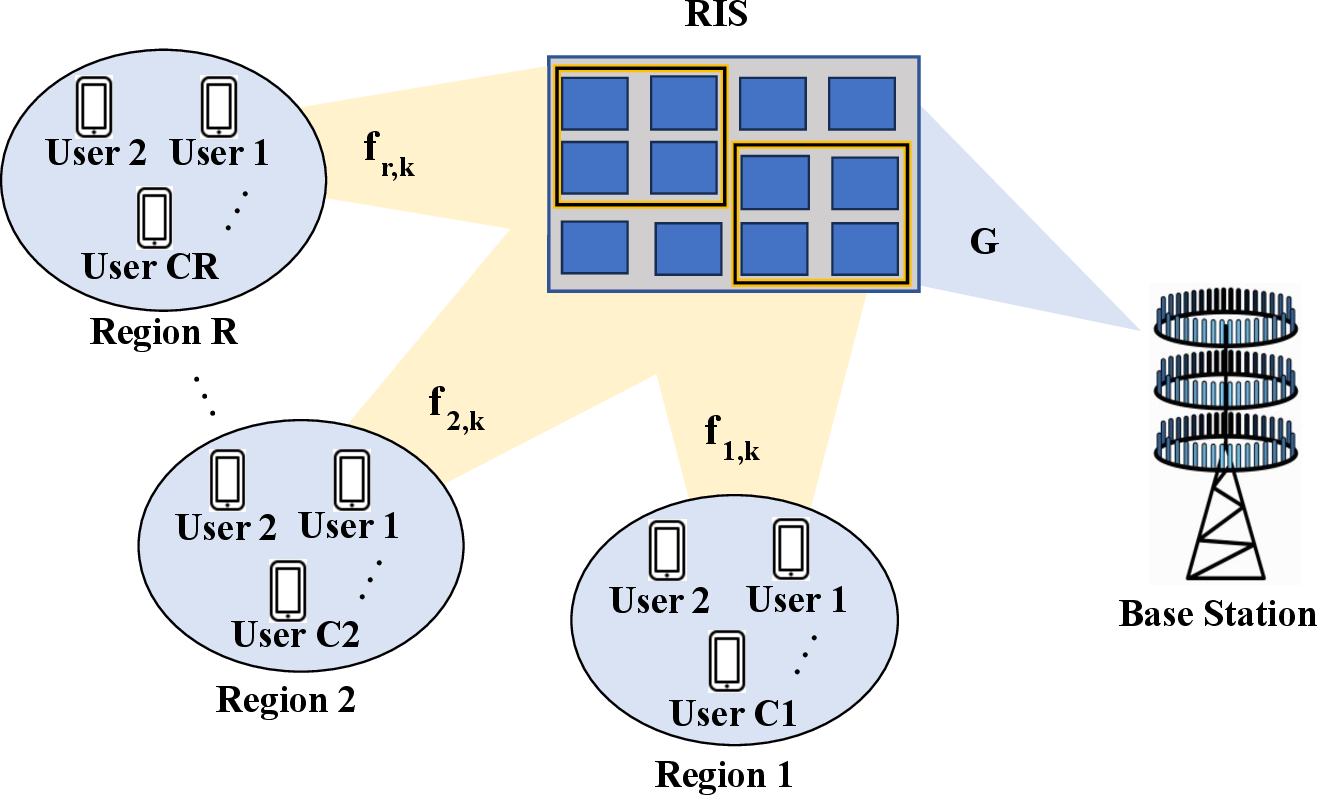}
\caption{Illustration of the RIS-assisted communication system.}
\label{fig:RIS_system}
\end{figure}

We consider a single-cell downlink system assisted by a reconfigurable intelligent surface (RIS), as illustrated in Fig.~\ref{fig:RIS_system}. The base station (BS) employs a uniform planar array (UPA) with $M=M_1M_2$ antennas and serves multiple single-antenna users. An RIS comprising $N=N_1N_2$ passive reflecting elements is deployed to improve propagation conditions, particularly in scenarios prone to blockage or high path loss. The BS configures the RIS via a dedicated low-rate control link~\cite{b3,b5,b8}, enabling dynamic adjustment of the reflection phases.

To account for spatial heterogeneity in the propagation environment, the service area is partitioned into $R$ spatial regions, each representing distinct propagation scenarios (e.g., varying angular spreads, scatterer clusters, or blockage profiles). Let $C_r$ denote the number of users in region $r$, where $r\in\{1,\ldots,R\}$. Users within the same region experience statistically similar channels, facilitating region-specific modeling, whereas inter-region variations may be significant, motivating adaptive estimation strategies.

We adopt a frequency-division duplex (FDD) framework; hence, uplink and downlink channels are not reciprocal. The direct BS--user link, if present, can be estimated by deactivating the RIS~\cite{b3} and is not modeled here, as our focus is the cascaded BS--RIS--user channel, which is critical for leveraging the RIS's reflective capabilities.

\subsection{Signal Model}
\label{subsec:signal-model}

Let $\mathbf{G}\in\mathbb{C}^{N\times M}$ denote the BS--RIS channel and $\mathbf{f}_{r,k}\in\mathbb{C}^{N\times 1}$ the RIS--user channel of the $k$-th user in region $r$ ($k=1,\ldots,C_r$). The downlink received signal at this user is
\begin{equation}
y_{r,k}=\mathbf{f}_{r,k}^{\mathrm H}\boldsymbol{\Phi}\,\mathbf{G}\,\mathbf{w}_{r,k}\,x_{r,k}+n_{r,k},
\label{eq:rx-signal}
\end{equation}
where $x_{r,k}\in\mathbb{C}$ is a unit-power symbol, $\mathbf{w}_{r,k}\in\mathbb{C}^{M\times 1}$ is the BS precoder, $n_{r,k}\sim\mathcal{CN}(0,\sigma_n^2)$ is additive white Gaussian noise (AWGN), and $\boldsymbol{\Phi}=\operatorname{diag}(\boldsymbol{\phi})$ with $\boldsymbol{\phi}=[\phi_1,\ldots,\phi_N]^{\mathrm T}$, $|\phi_n|=1$, is the RIS phase-shift matrix that controls the reflection phases.

We adopt the Saleh--Valenzuela geometry-based model~\cite{b4,b7} to capture the sparse multipath structure typical in mmWave and THz bands. The BS--RIS channel is
\begin{equation}
\mathbf{G}=\sqrt{\tfrac{MN}{L_G}}\sum_{l=1}^{L_G}\alpha_l^{G}\,
\mathbf{a}(\vartheta_l^{\mathrm{AoA}},\psi_l^{\mathrm{AoA}})\,
\mathbf{b}(\vartheta_l^{\mathrm{AoD}},\psi_l^{\mathrm{AoD}})^{\mathrm H},
\label{eq:G}
\end{equation}
where $L_G$ is the number of paths, $\alpha_l^G\sim\mathcal{CN}(0,1)$ represents the complex gain of the $l$-th path, $(\vartheta_l^{\mathrm{AoA}},\psi_l^{\mathrm{AoA}})$ are the azimuth and elevation angles of arrival at the RIS, and $(\vartheta_l^{\mathrm{AoD}},\psi_l^{\mathrm{AoD}})$ are the azimuth and elevation angles of departure from the BS. Owing to the fixed locations of the BS and RIS, $\mathbf{G}$ is quasi-static across users, allowing it to be estimated once and reused.

The RIS--user channel is
\begin{equation}
\mathbf{f}_{r,k}=\sqrt{\tfrac{N}{L_{r,k}}}\sum_{l=1}^{L_{r,k}}\alpha_l^{r,k}\,
\mathbf{a}(\vartheta_l^{r,k},\psi_l^{r,k}),
\label{eq:f}
\end{equation}
where $L_{r,k}$ is the number of paths, $\alpha_l^{r,k}\sim\mathcal{CN}(0,1)$, and $(\vartheta_l^{r,k},\psi_l^{r,k})$ are the departure angles from the RIS toward the user, reflecting the user's position-dependent scattering.

The UPA steering vectors $\mathbf{a}(\vartheta,\psi)\in\mathbb{C}^{N\times 1}$ and $\mathbf{b}(\vartheta,\psi)\in\mathbb{C}^{M\times 1}$ are
\begin{equation}
\begin{aligned}
\mathbf{a}(\vartheta,\psi) &=
\frac{1}{\sqrt{N}}
\Big[1, e^{-j \tfrac{2\pi d}{\lambda}\sin\psi\cos\vartheta}, \ldots, \\
&\quad e^{-j \tfrac{2\pi d}{\lambda}(N_2-1)\sin\psi\cos\vartheta}\Big]^{\mathrm{T}}
\otimes
\Big[1, e^{-j \tfrac{2\pi d}{\lambda}\cos\psi}, \ldots, \\
&\quad e^{-j \tfrac{2\pi d}{\lambda}(N_1-1)\cos\psi}\Big]^{\mathrm{T}},
\end{aligned}
\label{eq:array}
\end{equation}
with wavelength $\lambda$ and inter-element spacing $d=\lambda/2$; $\mathbf{b}(\cdot)$ is defined analogously for the BS UPA. This Kronecker structure captures the 2D array geometry, enabling angular resolution.

\subsection{Cascaded Channel Representation}
\label{subsec:cascaded}

Using~\eqref{eq:rx-signal}, define the cascaded BS--RIS--user channel
\begin{equation}
\mathbf{H}_{r,k}\triangleq \operatorname{diag}(\mathbf{f}_{r,k}^{\mathrm H})\,\mathbf{G}\in\mathbb{C}^{N\times M},
\end{equation}
so that
\begin{equation}
y_{r,k}=\boldsymbol{\phi}^{\mathrm H}\mathbf{H}_{r,k}\mathbf{w}_{r,k}\,x_{r,k}+n_{r,k}.
\label{eq:cascaded}
\end{equation}
This representation embeds the RIS phases into the effective channel, simplifying beamforming and estimation tasks.

\subsection{Pilot-Based Observation Model}
\label{subsec:pilot-model}

To acquire CSI, the BS transmits pilots across $Q$ time slots with distinct RIS configurations $\{\boldsymbol{\phi}_q\}_{q=1}^Q$. In slot $q$,
\begin{equation}
y_{r,k,q}^{\mathrm p}=\boldsymbol{\phi}_q^{\mathrm H}\mathbf{H}_{r,k}\mathbf{w}_{r,k}+n_{r,k,q},\quad
n_{r,k,q}\sim\mathcal{CN}(0,\sigma_n^2).
\end{equation}
Stacking over $Q$ slots gives
\begin{equation}
\mathbf{y}_{r,k}^{\mathrm p}
=\big(\mathbf{w}_{r,k}^{\mathrm T}\otimes \boldsymbol{\Theta}\big)\,\mathbf{h}_{r,k}+\mathbf{n}_{r,k},
\label{eq:pilot-model}
\end{equation}
where $\mathbf{y}_{r,k}^{\mathrm p}=[y_{r,k,1}^{\mathrm p},\ldots,y_{r,k,Q}^{\mathrm p}]^{\mathrm T}$,
$\boldsymbol{\Theta}=[\boldsymbol{\phi}_1,\ldots,\boldsymbol{\phi}_Q]^{\mathrm T}\in\mathbb{C}^{Q\times N}$,
$\mathbf{h}_{r,k}\triangleq \mathrm{vec}(\mathbf{H}_{r,k})\in\mathbb{C}^{NM\times 1}$,
and $\mathbf{n}_{r,k}$ collects noise terms. Define the measurement matrix
\begin{equation}
\boldsymbol{\Psi}_{r,k}\triangleq \mathbf{w}_{r,k}^{\mathrm T}\otimes \boldsymbol{\Theta}\in\mathbb{C}^{Q\times NM}.
\label{eq:psi}
\end{equation}
This linear model frames channel estimation as recovering a high-dimensional vector from low-dimensional noisy measurements, highlighting the compressive sensing aspect when $Q \ll NM$.

\subsection{RIS Grouping Model}
\label{subsec:grouping}

{\color{black}To reduce pilot overhead and computational complexity, we incorporate RIS element grouping via a fixed selection/averaging matrix $\mathbf{S}\in\{0,1\}^{N'\times N}$ that maps $N$ physical elements to $N'$ control units ($N'=N/g$ for cluster size $g$). This grouping aggregates nearby elements into clusters, preserving total reflection energy while reducing the degrees of freedom~\cite{bb1}. The RIS phase vectors satisfy
\begin{equation}
\boldsymbol{\phi}=\frac{1}{\sqrt{g}}\mathbf{S}^{\mathrm T}\bar{\boldsymbol{\phi}},\qquad \bar{\boldsymbol{\phi}}\in\mathbb{C}^{N'\times 1},\ \ |\bar{\phi}_i|=1,
\label{eq:grouping}
\end{equation}
and the effective cascaded channel is
\begin{equation}
\bar{\mathbf{H}}_{r,k}\triangleq \mathbf{S}\mathbf{H}_{r,k}\in\mathbb{C}^{N'\times M}.
\end{equation}
Let $\bar{\boldsymbol{\Theta}}=[\bar{\boldsymbol{\phi}}_1,\ldots,\bar{\boldsymbol{\phi}}_Q]^{\mathrm T}\in\mathbb{C}^{Q\times N'}$. The stacked pilot model becomes
\begin{equation}
\mathbf{y}_{r,k}^{\mathrm p}
=\big(\mathbf{w}_{r,k}^{\mathrm T}\otimes \bar{\boldsymbol{\Theta}}\big)\,\bar{\mathbf{h}}_{r,k}+\mathbf{n}_{r,k},
\qquad
\bar{\mathbf{h}}_{r,k}\triangleq \mathrm{vec}(\bar{\mathbf{H}}_{r,k}).
\label{eq:pilot-model-grouped}
\end{equation}
Define $\bar{\boldsymbol{\Psi}}_{r,k}\triangleq \mathbf{w}_{r,k}^{\mathrm T}\otimes \bar{\boldsymbol{\Theta}}\in\mathbb{C}^{Q\times N'M}$. Grouping thus compresses the estimation problem, enabling operation with fewer pilots at the cost of some resolution loss.

Throughout this work we model the RIS as having ideal continuous-valued phase control. Each element uses $\phi_{q,n}=e^{j\theta_{q,n}}$ with $\theta_{q,n}\in[0,2\pi)$ and $|\phi_{q,n}|=1$. This corresponds to an upper bound on the performance of practical RISs. Practical RISs often employ $B$-bit phase shifters with $\theta_{q,n}$ restricted to a finite set~\cite{bb2}. Explicit few-bit quantization constraints on $\{\theta_{q,n}\}$ into the pilot and beamforming design are left for future work. The linear observation model in~\eqref{eq:pilot-model} remains valid when $\phi_{q,n}$ is replaced by its quantized counterpart. The grouped variant in~\eqref{eq:pilot-model-grouped} also remains valid under quantization.}

\subsection{Problem Formulation}
\label{sec:problem}

\textbf{Goal:} For each user $(r,k)$, estimate the cascaded channel vector from noisy pilot observations:
\begin{equation}
\underbrace{\mathbf{y}_{r,k}^{\mathrm p}}_{\in\mathbb{C}^{Q}}
=
\underbrace{\boldsymbol{\Psi}_{r,k}}_{\in\mathbb{C}^{Q\times NM}}
\underbrace{\mathbf{h}_{r,k}}_{\in\mathbb{C}^{NM}}
+\ \mathbf{n}_{r,k}.
\label{eq:observation}
\end{equation}

\textbf{Estimator class:} Let $g_{\boldsymbol{\vartheta}}(\cdot)$ denote an estimator with parameters $\boldsymbol{\vartheta}$ (e.g., a deep neural network). The channel estimate is
\begin{equation}
\hat{\mathbf{h}}_{r,k}=g_{\boldsymbol{\vartheta}}\big(\mathbf{y}_{r,k}^{\mathrm p};\boldsymbol{\Psi}_{r,k}\big).
\label{eq:estimator}
\end{equation}
The estimator may condition on the known measurement matrix to incorporate pilot information.

\textbf{Performance metric:} We use normalized mean-squared error (NMSE):
\begin{equation}
\mathrm{NMSE}\triangleq
\mathbb{E}\!\left[\frac{\|\hat{\mathbf{h}}_{r,k}-\mathbf{h}_{r,k}\|_2^2}{\|\mathbf{h}_{r,k}\|_2^2}\right].
\label{eq:nmse}
\end{equation}
This metric normalizes reconstruction error by channel energy, providing a scale-invariant measure of accuracy.

\textbf{Fixed-pilot estimation problem (PF-1):} Given pilot matrix $\boldsymbol{\Theta}$ (or $\bar{\boldsymbol{\Theta}}$) and beamformer $\mathbf{w}_{r,k}$,
\begin{equation}
\boxed{\
\begin{aligned}
& \min_{\boldsymbol{\vartheta}}\
\mathbb{E}\big[\mathrm{NMSE}\big] \\
& \text{s.t.}\ \ Q\ll NM\ \ (\text{or }Q\ll N'M),\ |\phi_n|=1.\
\end{aligned}
}
\label{eq:pf1}
\end{equation}
The expectation is over channel, noise, users $(r,k)$, and regions, ensuring robustness across diverse scenarios.

\textbf{Joint pilot-estimator design (PF-2):} If pilot patterns are optimizable,
\begin{equation}
\boxed{\
\begin{aligned}
& \min_{\boldsymbol{\vartheta},\ \boldsymbol{\Theta}\ \text{(or }\bar{\boldsymbol{\Theta}}\text{)}}\
\mathbb{E}\big[\mathrm{NMSE}\big] \\
& \text{s.t.}\ |\phi_{q,n}|=1,  Q\le Q_{\max},\, \text{PAPR}\footnote{peak-to-average-power ratio}.\
\end{aligned}
}
\label{eq:pf2}
\end{equation}
In this work we adopt fixed, unit-modulus pilots with entries on the unit circle
(e.g., $\phi_{q,n}\in\{+1,-1\}$ or $\phi_{q,n}=e^{j\theta_{q,n}}$). If needed, energy normalization is applied to the stacked model (e.g., using $\tilde{\mathbf{y}}_{r,k}^{\mathrm p}=\mathbf{y}_{r,k}^{\mathrm p}/\sqrt{Q}$), without violating $|\phi_{q,n}|=1$.

\textbf{Baselines:} For completeness, when $Q\ge NM$ (or $Q\ge N'M$) the LS estimator reads
\begin{equation}
\hat{\mathbf{h}}_{r,k}^{\mathrm{LS}}=
\big(\boldsymbol{\Psi}_{r,k}^{\mathrm H}\boldsymbol{\Psi}_{r,k}\big)^{-1}\boldsymbol{\Psi}_{r,k}^{\mathrm H}\mathbf{y}_{r,k}^{\mathrm p}.
\label{eq:ls}
\end{equation}
The LS solution exists if $\mathrm{rank}(\boldsymbol{\Psi}_{r,k})=NM$ (or $N'M$);
otherwise we use the Moore–Penrose pseudoinverse.

MMSE requires a channel covariance model; we adopt a data-driven covariance estimated from samples in the baselines section ( Sec.~\ref{sec:simulation}).

\textbf{Identifiability and pilot design:} Identifiability requires $\mathrm{rank}(\boldsymbol{\Psi}_{r,k})\ge \mathrm{dim}(\mathrm{span}(\mathbf{h}_{r,k}))$. With grouping, the intrinsic dimension reduces from $NM$ to $N'M$, permitting $Q\ll NM$. Random unit-modulus pilot patterns with entries in $\{\pm 1/\sqrt{Q}\}$ ensure well-conditioned sensing matrices with high probability, as supported by standard results on random design matrices~\cite{b16}.

\vspace{2em}

\section{Proposed DML-Based Neural Network}
\label{sec:proposed-dml}

In this section, we introduce a distributed machine learning (DML) framework for estimating the cascaded channel based on the observation models presented in~\eqref{eq:pilot-model} and~\eqref{eq:pilot-model-grouped}. This design is tailored to achieve three primary objectives: (i) precise reconstruction of the channel from a limited number of pilots ($Q \ll NM$ or $Q \ll N'M$), (ii) robust performance across diverse spatial regions as defined in Sec.~\ref{sec:system-model}, and (iii) minimal computational complexity during online inference at the UE. By leveraging DML, the framework enables collaborative training on decentralized, non-IID datasets without requiring the exchange of raw channel data, thereby preserving user privacy and reducing communication overhead. This approach builds upon the federated learning paradigm to address the generalization challenges inherent in heterogeneous propagation environments.

\subsection{Input Encoding and Output Reconstruction}
\label{subsec:io}

For a given user $(r,k)$, the $Q$ complex-valued pilot observations are aggregated into the vector $\mathbf{y}_{r,k}^{\mathrm{p}} \in \mathbb{C}^{Q}$. To facilitate processing by convolutional neural networks (CNNs), which are effective for capturing spatial and structural patterns in data, this vector is reshaped into a two-channel real-valued tensor
\begin{equation}
\mathbf{Y}_{r,k} \in \mathbb{R}^{Q_1 \times Q_2 \times 2}, \quad Q = Q_1 Q_2,
\label{eq:pilot-tensor}
\end{equation}
where the two channels correspond to the real and imaginary components of the pilots. This tensor representation exploits the grid-like structure of the observations, enabling the network to learn local correlations efficiently.

The estimator $g_{\boldsymbol{\vartheta}}(\cdot)$ then produces the estimated cascaded channel vector, either $\hat{\mathbf{h}}_{r,k} \in \mathbb{C}^{NM}$ in the ungrouped case or $\hat{\bar{\mathbf{h}}}_{r,k} \in \mathbb{C}^{N'M}$ when grouping is applied:
\begin{equation}
\hat{\mathbf{h}}_{r,k} = g_{\boldsymbol{\vartheta}}(\mathbf{Y}_{r,k}) \quad \text{or} \quad \hat{\bar{\mathbf{h}}}_{r,k} = g_{\boldsymbol{\vartheta}}(\mathbf{Y}_{r,k}),
\label{eq:estimator-impl}
\end{equation}
as an implementation of the general estimator outlined in~\eqref{eq:estimator}. The complex-valued output is handled by generating two real-valued channels, which are subsequently combined to form the final estimate. This encoding strategy ensures compatibility with standard deep learning architectures while preserving the intrinsic properties of the wireless channel data.

\subsection{Architecture: Region-Gated Mixture of Experts}
\label{subsec:arch}

The proposed estimator adopts a region-gated mixture-of-experts (MoE) architecture, which is particularly suited for handling non-IID data distributions across spatial regions. In this setup, the MoE consists of three interconnected modules: a scenario classifier acting as the gate, a set of region-specific feature extractors serving as experts, and a feature mapper. This structure draws inspiration from sparsely-gated MoE models~\cite{b18}, where the gate dynamically selects the most appropriate expert based on the input, thereby enhancing specialization and efficiency. By activating only one expert per inference, the architecture maintains low latency while adapting to regional variations in channel characteristics, such as differing angular spreads or blockage profiles.

\subsubsection{Scenario Classifier (Gating Network)}
\label{subsubsec:classifier}

\begin{figure}[htbp]
\centering
\includegraphics[width=0.8\linewidth]{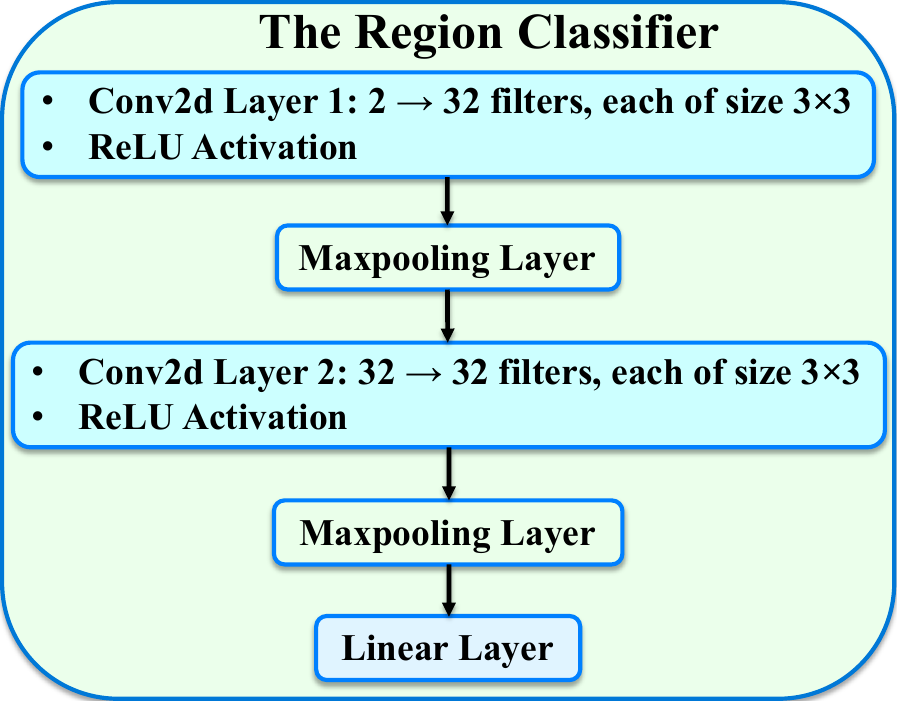}
\caption{The region classifier.}
\label{fig:region_classifier}
\end{figure}

The gating network is implemented as a lightweight CNN, denoted $C_{\boldsymbol{\phi}}: \mathbb{R}^{Q_1 \times Q_2 \times 2} \to \Delta^{R-1}$, where $\Delta^{R-1} \triangleq \{\mathbf{p} \in \mathbb{R}_{+}^{R} : \sum_i p^{(i)} = 1\}$ represents the $(R-1)$-simplex. This module processes the input tensor $\mathbf{Y}_{r,k}$ to produce a probability distribution $\mathbf{p}_{r,k} = [p^{(1)}, \dots, p^{(R)}]$, with $\sum_i p^{(i)} = 1$, over the $R$ regions. The predicted region index is then determined as $\hat{r} = \arg\max_i p^{(i)}$. This gating mechanism embodies the sparse routing principle of MoE~\cite{b18}, ensuring computational efficiency by directing the input to only the most relevant expert. The architecture of the classifier is illustrated in Fig.~\ref{fig:region_classifier}, which typically includes a few convolutional layers followed by a softmax output to generate the probabilities.

\subsubsection{Region Experts (Feature Extractors)}
\label{subsubsec:experts}

To capture region-specific channel features, we deploy $R$ independent experts $\{E_{\boldsymbol{\theta}_1}, \dots, E_{\boldsymbol{\theta}_R}\}$. During inference, only the expert corresponding to the predicted region, $E_{\boldsymbol{\theta}_{\hat{r}}}$, is activated, minimizing unnecessary computations. Each expert is structured as a sequence of three $3 \times 3$ convolutional layers, each with 32 output channels, interleaved with batch normalization (BN) and rectified linear unit (ReLU) activations. Same-padding is employed to preserve the spatial dimensions throughout the network. The input to the first layer has 2 channels (real and imaginary parts), expanding to 32 channels, which are maintained in subsequent layers.

\begin{figure}[htbp]
\centering
\includegraphics[width=0.80\linewidth]{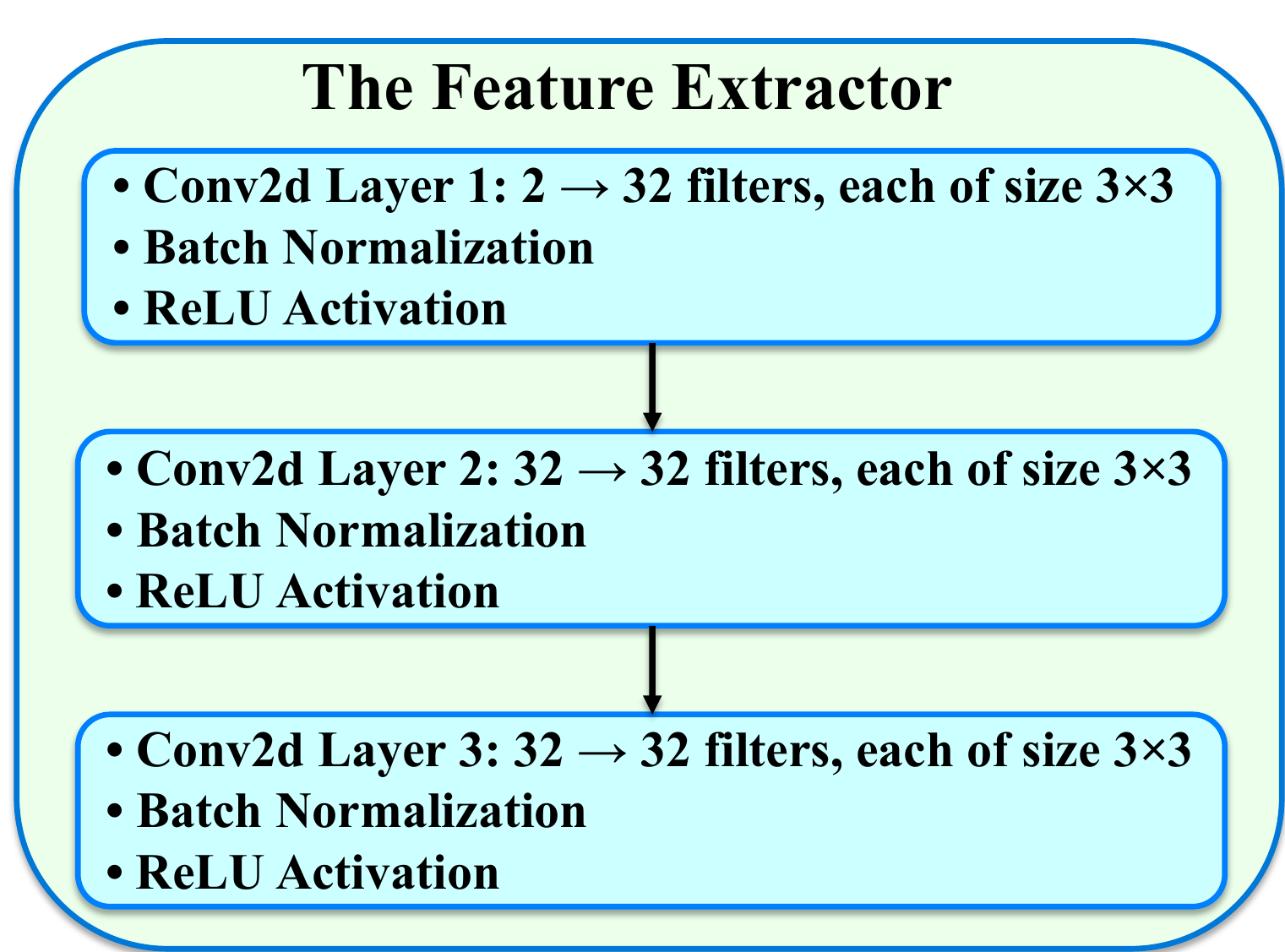}
\caption{The feature extractor.}
\label{fig:feature_extractor}
\end{figure}

The feature extraction process is formalized as
\begin{equation}
\mathbf{Z}_{r,k} = E_{\boldsymbol{\theta}_{\hat{r}}}(\mathbf{Y}_{r,k}),
\label{eq:expert}
\end{equation}
yielding a region-tailored feature map $\mathbf{Z}_{r,k}$. This design leverages the convolutional operations to exploit local dependencies in the pilot tensor, such as spatial correlations induced by the RIS geometry. The expert architecture is depicted in Fig.~\ref{fig:feature_extractor}, highlighting the three-layer stack (Conv $3\times3$ $\to$ BN $\to$ ReLU) that processes the input to produce discriminative features adapted to the specific propagation scenario.

\subsubsection{Feature Mapper}
\label{subsubsec:mapper}

\begin{figure}[htbp]
\centering
\includegraphics[width=0.45\linewidth]{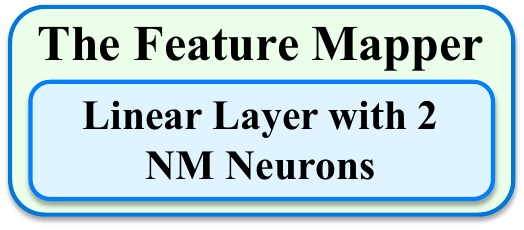}
\caption{The feature mapper.}
\label{fig:feature_mapper}
\end{figure}

The final module, the feature mapper $M_{\boldsymbol{\psi}}$, transforms the extracted features into the estimated channel vector through a linear projection. This yields
\begin{equation}
\hat{\mathbf{h}}_{r,k} = M_{\boldsymbol{\psi}}(\mathbf{Z}_{r,k}),
\label{eq:mapper}
\end{equation}
depending on whether grouping is utilized. In implementation, the mapper can be realized via a $1 \times 1$ convolution for channel-wise mixing, followed by flattening and a fully connected layer to produce the output of dimension $2NM$ (or $2N'M$) real values, which are then reassembled into the complex channel estimate. This linear mapping ensures that the high-level features from the expert are directly translated into the channel domain, facilitating end-to-end learning. The mapper's structure is shown in Fig.~\ref{fig:feature_mapper}.

\paragraph*{Soft gating:} For enhanced robustness during training, a soft-gated variant computes a weighted combination of expert outputs as $\tilde{\mathbf{Z}}_{r,k} = \sum_{i=1}^{R} p_{r,k}^{(i)} E_{\boldsymbol{\theta}_i}(\mathbf{Y}_{r,k})$ before applying $M_{\boldsymbol{\psi}}$, aligning with the probabilistic routing in~\cite{b18}. However, we primarily employ hard gating at inference to optimize computational efficiency.

\subsection{Learning Objectives}
\label{subsec:objectives}

The training process optimizes two distinct loss functions to separately refine the classifier and the estimation components, ensuring stable convergence in the presence of non-IID data.

\textbf{Estimation loss:} The primary objective is to minimize the empirical normalized mean-squared error (NMSE) over mini-batches, as defined in~\eqref{eq:nmse}:
\begin{equation}
\mathcal{L}_{\mathrm{est}}(\boldsymbol{\vartheta}) = \frac{1}{|\mathcal{B}|} \sum_{(r,k) \in \mathcal{B}} \frac{\|\hat{\mathbf{h}}_{r,k} - \mathbf{h}_{r,k}\|_2^2}{\|\mathbf{h}_{r,k}\|_2^2} \quad \text{or} \quad \frac{\|\hat{\bar{\mathbf{h}}}_{r,k} - \bar{\mathbf{h}}_{r,k}\|_2^2}{\|\bar{\mathbf{h}}_{r,k}\|_2^2}.
\label{eq:emp-nmse}
\end{equation}
This loss directly targets channel reconstruction accuracy, normalized to account for varying channel strengths across samples.

\textbf{Classification loss:} To train the gating network, we use cross-entropy loss against the true region labels:
\begin{equation}
\mathcal{L}_{\mathrm{cls}}(\boldsymbol{\phi}) = -\frac{1}{|\mathcal{B}|} \sum_{(r,k) \in \mathcal{B}} \sum_{i=1}^{R} \mathbf{1}\{i = r\} \log p_{r,k}^{(i)}.
\label{eq:cls}
\end{equation}
We adopt a two-stage training strategy: first, pretrain the classifier $C_{\boldsymbol{\phi}}$ using $\mathcal{L}_{\mathrm{cls}}$ to achieve reliable region identification; then, fix the classifier and optimize the experts and mapper jointly with $\mathcal{L}_{\mathrm{est}}$. This separation mitigates instability arising from joint optimization in heterogeneous settings and ensures that the gate accurately routes inputs to appropriate experts.

\subsection{Distributed Training Protocol}
\label{subsec:dml}

The DML training employs a synchronous federated learning protocol, with the BS serving as the central parameter server and the UEs as clients. This setup leverages the synchronous Federated Averaging (FedAvg) algorithm~\cite{b17} to aggregate local updates, effectively handling the non-IID nature of channel data across regions. Let $\boldsymbol{\vartheta} = (\{\boldsymbol{\theta}_r\}_{r=1}^R, \boldsymbol{\psi})$ denote the trainable parameters of the experts and mapper. In each communication round, the BS broadcasts the current model, clients perform local gradient computations on their private datasets, and the BS aggregates these updates using data-weighted averaging to update the global model.

The detailed procedure is outlined in Algorithm~\ref{alg:dml-train}. This protocol not only improves generalization by incorporating diverse regional data but also minimizes communication costs, as only model updates (or gradients) are exchanged, not raw data.

\begin{algorithm}[htbp]
\caption{DML Training of the Proposed Estimator}
\label{alg:dml-train}

{\color{black}%
\begin{algorithmic}[1]
\Require Client datasets $\{\mathcal{D}_{r,k}\}$ with $(\mathbf{Y}_{r,k},\mathbf{h}_{r,k})$ or $(\mathbf{Y}_{r,k},\bar{\mathbf{h}}_{r,k})$; pretrained $C_{\boldsymbol{\phi}}$ via~\eqref{eq:cls}; rounds $T$; stepsizes $\eta^{(t)}$.
\Ensure Trained parameters $\boldsymbol{\vartheta}$.
\State Initialize $\boldsymbol{\vartheta}^{(0)}$.
\For{$t=0$ to $T-1$}
  \State Broadcast $\boldsymbol{\vartheta}^{(t)}$ from BS to clients.
  \For{each client $(r,k)$ in parallel}
    \State Sample mini-batches $\mathcal{B}\subset\mathcal{D}_{r,k}$.
    \State For $(\mathbf{Y}_{r,k},\cdot)\!\in\!\mathcal{B}$, compute $\hat r=\arg\max C_{\boldsymbol{\phi}}(\mathbf{Y}_{r,k})$; forward through $E_{\boldsymbol{\theta}_{\hat r}}$ and $M_{\boldsymbol{\psi}}$; accumulate~\eqref{eq:emp-nmse}.
    \State Local gradient: $\mathbf{g}_{r,k}^{(t)}=\nabla_{\boldsymbol{\vartheta}}\mathcal{L}_{\mathrm{est}}(\boldsymbol{\vartheta}^{(t)};\mathcal{B})$ (nonzero only for $E_{\boldsymbol{\theta}_{\hat r}}$, $M_{\boldsymbol{\psi}}$).
    \State Upload $\mathbf{g}_{r,k}^{(t)}$ (or model delta) to BS.
  \EndFor
  \State Aggregate at BS with weights $p_{r,k}=|\mathcal{D}_{r,k}|/\sum_{r',k'}|\mathcal{D}_{r',k'}|$:
  \begin{equation}
    \boldsymbol{\vartheta}^{(t+1)}=\boldsymbol{\vartheta}^{(t)}-\eta^{(t)}\sum_{r,k}p_{r,k}\,\mathbf{g}_{r,k}^{(t)}.
    \label{eq:agg}
  \end{equation}
\EndFor
\end{algorithmic}
}%
\end{algorithm}

\paragraph*{Communication efficiency:} Extensions such as gradient compression, client subsampling, and secure aggregation can be integrated to further enhance efficiency, consistent with advancements in federated learning~\cite{b17}.

\subsection{Inference/Evaluation}
\label{subsec:inference}

During inference, the UE executes a streamlined forward pass: the classifier identifies the region, the corresponding expert extracts features, and the mapper reconstructs the channel. This process requires no further optimization and is computationally lightweight due to the single-expert activation. Algorithm~\ref{alg:dml-infer} details the evaluation pipeline, which computes NMSE across SNR levels and benchmarks against classical methods like LS and MMSE under identical pilot constraints.

\begin{algorithm}[htbp]
\caption{Inference/Evaluation Pipeline}
\label{alg:dml-infer}
\begin{algorithmic}[1]
\Require Trained $\boldsymbol{\vartheta}$ and $C_{\boldsymbol{\phi}}$; test set $\{(\mathbf{Y}_{r,k},\mathbf{h}_{r,k})\}$ or $\{(\mathbf{Y}_{r,k},\bar{\mathbf{h}}_{r,k})\}$.
\Ensure NMSE (~\eqref{eq:nmse}) vs. SNR.
\For{each SNR point}
  \For{each $(\mathbf{Y}_{r,k},\cdot)$}
    \State $\mathbf{p}_{r,k}=C_{\boldsymbol{\phi}}(\mathbf{Y}_{r,k})$, $\hat r=\arg\max_i p_{r,k}^{(i)}$.
    \State $\hat{\mathbf{h}}_{r,k}$ (or $\hat{\bar{\mathbf{h}}}_{r,k}$) via~\eqref{eq:mapper} using $E_{\boldsymbol{\theta}_{\hat r}}$, $M_{\boldsymbol{\psi}}$.
  \EndFor
  \State Compute NMSE using~\eqref{eq:nmse}; compare to LS/MMSE with the same pilots.
\EndFor
\end{algorithmic}
\end{algorithm}

\subsection{Computational Complexity}
\label{subsec:complexity}

The complexity analysis focuses on multiply-accumulate (MAC) operations, providing a theoretical measure of efficiency. For the active expert with three $3 \times 3$ convolutional layers and 32 channels each (preserving spatial dimensions $E_1 \times E_2 = Q_1 \times Q_2$), the convolutional MAC count is
\begin{equation}
\mathcal{C}_{\mathrm{conv}} = \sum_{\ell=1}^{3} E_1 E_2 \cdot 9 \cdot C_{\mathrm{in}}^{(\ell)} C_{\mathrm{out}}^{(\ell)} = 9 E_1 E_2 (2 \cdot 32 + 32 \cdot 32 + 32 \cdot 32),
\label{eq:conv-complexity}
\end{equation}
accounting for the input channels (2 for the first layer, 32 thereafter). {\color{blue}The mapper's complexity is
\begin{equation}
\mathcal{C}_{\mathrm{map}} = \mathcal{O}\big(E_1 E_2 \cdot 32 \cdot (2NM)\big),
\label{eq:map-complexity}
\end{equation}
reflecting the linear projection to the channel space. }The classifier's overhead is negligible compared to these terms. By gating to a single expert, the overall runtime remains comparable to a compact CNN, as validated in sparsely-gated MoE literature~\cite{b18}, making the framework suitable for resource-constrained UEs.

\vspace{2em}

\begin{table}[t]
\caption{Simulation Parameters and Configurations}
\label{tab:sim_params}
\centering
\begin{threeparttable}
\setlength{\tabcolsep}{3.5pt}
\renewcommand{\arraystretch}{1.05}
\footnotesize
\begin{tabular}{@{}p{0.47\columnwidth} p{0.47\columnwidth}@{}}
\toprule
\hline
\multicolumn{2}{@{}l}{\textbf{A. Arrays, Pilots, and SNR}}\\
\midrule
BS array & UPA $M_1{\times}M_2=4{\times}4$ ($M=16$) \\
RIS array & UPA $N_1{\times}N_2=8{\times}8$ ($N=64$) \\
Carrier / spacing & $f_c=28$\,GHz; $\lambda{=}c/f_c$; $d=\lambda/2$ \\
RIS grouping & Cluster size $g=4$; effective $N'=N/g=16$ \\
Pilot budget & $Q=32$ with stacking $Q_1{\times}Q_2=8{\times}4$ \\
RIS pilots $\boldsymbol{\Theta}$ & unit-modulus entries $\phi_{q,n}=e^{j\theta_{q,n}}$, $\theta_{q,n}\sim\mathrm{Unif}[0,2\pi)$\footnotemark \\
BS pilot beamformer $\mathbf{w}_{r,k}$ & Random unit-norm (fixed during pilots) \\
Noise \& SNR & $n\!\sim\!\mathcal{CN}(0,\sigma_n^2)$; $\mathrm{SNR}=1/\sigma_n^2$ (linear) \\
\midrule
\multicolumn{2}{@{}l}{\textbf{B. Channel Models (\eqref{eq:G}, \eqref{eq:f})}}\\
\midrule
BS–RIS channel $\mathbf{G}$ & Saleh–Valenzuela; $L_G=3$; $\alpha_\ell^{G}\!\sim\!\mathcal{CN}(0,1)$ \\
Angles for $\mathbf{G}$ & AoA/AoD az.,el. $\sim \mathrm{Unif}(-\pi/2,\pi/2)$ \\
RIS–user $\mathbf{f}_{r,k}$ & Saleh–Valenzuela; $L_{r,k}=3$; $\alpha_\ell^{r,k}\!\sim\!\mathcal{CN}(0,1)$ \\
Region partitions (elev.) & $(-\pi/2,-\pi/6)$, $(-\pi/6,\pi/6)$, $(\pi/6,\pi/2)$ \\
Regions / users & $R=3$; users per region $C_r=3$ \\
\midrule
\multicolumn{2}{@{}l}{\textbf{C. Learning Setup (Sec.~\ref{sec:proposed-dml})}}\\
\midrule
Samples per user & $20{,}000$ (train/val: $90\%/10\%$) \\
Total DML samples & $3{\times}3{\times}20{,}000=180{,}000$ \\
Optimizer & Adam; batch size $256$ \\
Learning rate & $10^{-3}$; halved every $30$ epochs; $100$ epochs \\
Supervision (labels) & LS at $10$\,dB SNR (training targets) \\
Test set & $10{,}000$ samples; NMSE vs.\ SNR \\
\hline
\bottomrule
\end{tabular}
\begin{tablenotes}
  \footnotesize
\item{{\color{black}Unit-modulus pilots satisfy $|\phi_{q,n}|=1$. For energy normalization, we use
$\tilde{\mathbf{y}}^{\mathrm p}=\mathbf{y}^{\mathrm p}/\sqrt{Q}$
without altering~\eqref{eq:pilot-model}.}}
\end{tablenotes}
\end{threeparttable}
\end{table}

\section{Simulation Results and Discussion}
\label{sec:simulation}
In this section, we assess the performance of the proposed DML-based estimator using the observation models in~\eqref{eq:pilot-model} and~\eqref{eq:pilot-model-grouped}. We compare it against LS and MMSE baselines, focusing on the NMSE metric defined in~\eqref{eq:nmse} as a function of SNR. These evaluations demonstrate the estimator's ability to achieve high accuracy with reduced pilot overhead, robust generalization across regions, and low computational demands, validating its suitability for practical RIS-assisted systems.

\subsection{Simulation Setup}
\label{subsec:sim-setup}
The simulation parameters are summarized in Table~\ref{tab:sim_params}, categorized into array configurations, channel models, and learning setups for reproducibility and clarity.

\textbf{Array geometry and carrier:} The BS is equipped with a UPA of $M = M_1 \times M_2 = 4 \times 4 = 16$ antennas, while the RIS comprises $N = N_1 \times N_2 = 8 \times 8 = 64$ elements. The inter-element spacing is set to $d = \lambda/2$ at a carrier frequency of $f_c = 28$~GHz, promoting angular resolution in mmWave bands.

\textbf{Grouping:} To compress the channel dimensions, RIS elements are grouped into clusters of size $g=4$, resulting in $N' = N/g = 16$ effective control units. The grouping matrix $\mathbf{S} \in {0,1}^{N' \times N}$ assigns each control unit to exactly $g$ physical elements. This aggregation preserves total reflection energy while reducing the estimation complexity; an optional scaling of $1/\sqrt{g}$ in $\bar{\boldsymbol{\phi}}$ can maintain per-symbol energy if needed, though our results use receiver-side normalization and are robust to this choice. The grouped cascaded channel becomes $\bar{\mathbf{H}}{r,k} = \mathbf{S} \mathbf{H}{r,k}$, with the pilot model following~\eqref{eq:pilot-model-grouped} and sensing matrix $\bar{\boldsymbol{\Psi}}{r,k} = \mathbf{w}{r,k}^{\mathrm{T}} \otimes \bar{\boldsymbol{\Theta}}$.

\textbf{Pilots and BS precoder:} A fixed unit-norm precoding vector $\mathbf{w}_{r,k}$ is randomly selected from the complex unit sphere and held constant across pilots to simulate isotropic probing. The RIS phase configurations are unit-modulus, with entries drawn i.i.d. from ${\pm 1}$ (or uniformly on the unit circle), forming $\boldsymbol{\Theta} = [\boldsymbol{\phi}_1, \dots, \boldsymbol{\phi}_Q]^{\mathrm{T}}$. For grouped cases, $\bar{\boldsymbol{\Theta}} \in \mathbb{C}^{Q \times N'}$ uses similar patterns. We operate in the short-pilot regime with $Q=32$, reshaped as $Q_1 \times Q_2 = 8 \times 4$ for the input tensor in~\eqref{eq:pilot-tensor}.
\textbf{Channel model:} Both the BS-RIS channel $\mathbf{G}$ and RIS-user channel $\mathbf{f}_{r,k}$ follow the Saleh-Valenzuela model in~\eqref{eq:G} and~\eqref{eq:f}, respectively, each with 3 paths. Path gains are i.i.d. $\mathcal{CN}(0,1)$, and azimuth/elevation angles are uniformly distributed in $(-\pi/2, \pi/2)$, with elevation angles at the RIS constrained by region partitions to simulate heterogeneous propagation.

\textbf{Regions and datasets:} The coverage area is divided into $R=3$ elevation-based regions: $(-\pi/2, -\pi/6)$, $(-\pi/6, \pi/6)$, and $(\pi/6, \pi/2)$. Each region contains $C_r=3$ users, with 20,000 channel samples generated per user, yielding a total of 180,000 samples for DML (90\% training, 10\% validation). Training labels are LS estimates at 10 dB SNR to reflect realistic supervision; test evaluations use 10,000 independent samples per scenario, ensuring unbiased assessment.

\textbf{Learning protocol:} Training employs the Adam optimizer with a batch size of 256, an initial learning rate of $10^{-3}$ halved every 30 epochs over 100 epochs total. The DML process uses synchronous FedAvg~\cite{b17} with data-weighted aggregation. The classifier is pretrained via cross-entropy loss, then fixed during NMSE-based optimization of experts and mapper. Inference defaults to hard gating for efficiency.

\textbf{Baselines:} LS follows~\eqref{eq:ls}, using the Moore-Penrose pseudoinverse for underdetermined systems. MMSE incorporates a sample covariance matrix estimated from training data. To provide a fair comparison in the short-pilot regime, baselines are evaluated at higher pilot budgets ($Q=1024$ ungrouped, $Q=256$ grouped) where they achieve stability, highlighting the proposed method's overhead reduction.

\subsection{Main Results}
\label{subsec:main-results}

We present NMSE versus SNR curves to illustrate the estimator's performance, emphasizing gains from grouping, DML, and gating.

\subsubsection{Ungrouped RIS: Short-Pilot Learning vs. Long-Pilot Baselines}
\begin{figure}[htbp]
\centering
\includegraphics[width=\linewidth]{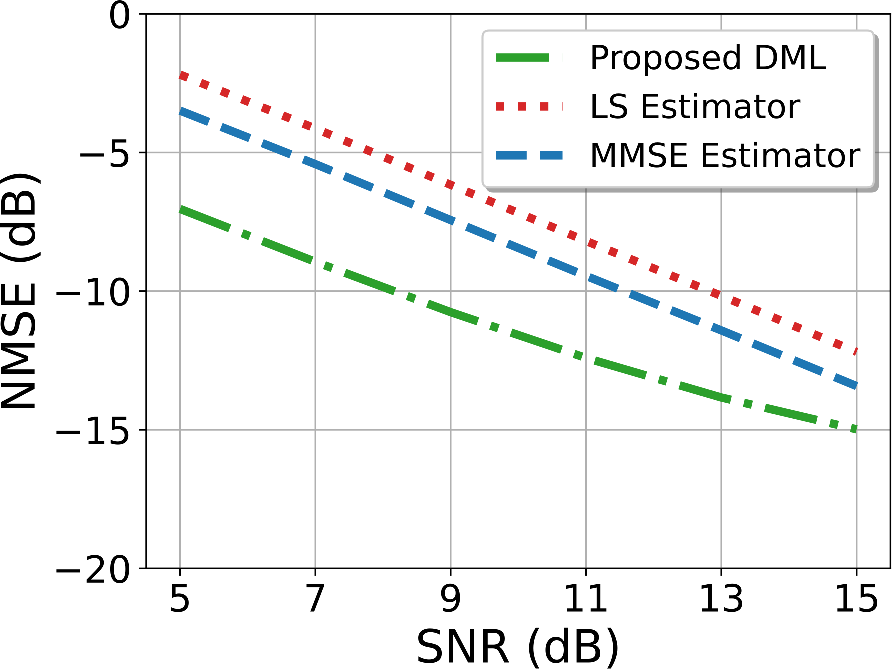}
\caption{NMSE with ungrouped RIS ($N=64$). Proposed model uses $Q=32$. Baselines use $Q=1024$.}
\label{fig:nmse_before_grouping}
\end{figure}

Fig.~\ref{fig:nmse_before_grouping} shows results without grouping ($N=64$). The proposed estimator uses $Q=32$ pilots, while LS and MMSE require $Q=1024$ for reliable inversion. Our method delivers superior NMSE across all SNRs, particularly in low-SNR regimes where baselines suffer from noise amplification. This advantage arises from the CNN's ability to exploit structural priors in the pilot tensor and the MoE's region-specific adaptation, enabling effective reconstruction despite compressive measurements.

\subsubsection{Effect of Training Only on a Single Region}
\begin{figure}[htbp]
\centering
\includegraphics[width=\linewidth]{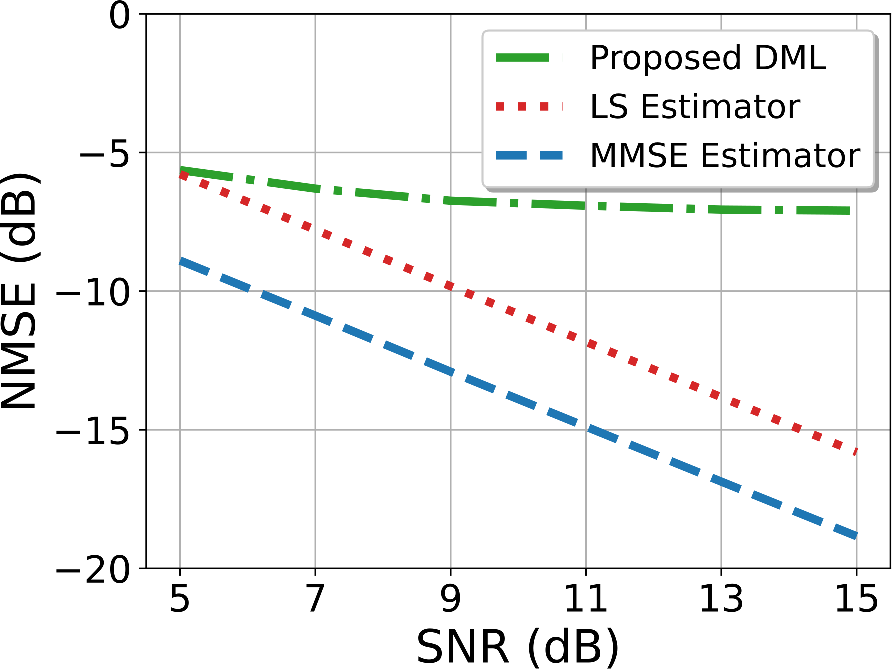}
\caption{NMSE after grouping ($N'=16$) when training on a single region only. Generalization across regions degrades.}
\label{fig:nmse_single_region}
\end{figure}

Fig.~\ref{fig:nmse_single_region} examines generalization when training is limited to one region (after grouping to $N'=16$). The estimator overfits to the training region's statistics, leading to degraded NMSE on mismatched test data. In contrast, LS and MMSE with ample pilots ($Q=256$) exhibit better cross-region stability but at higher overhead. This underscores the need for DML to aggregate diverse data and gating to handle heterogeneity.

\subsubsection{Grouped RIS: Accuracy, Robustness, and Pilot Reduction}
\begin{figure}[htbp]
\centering
\includegraphics[width=\linewidth]{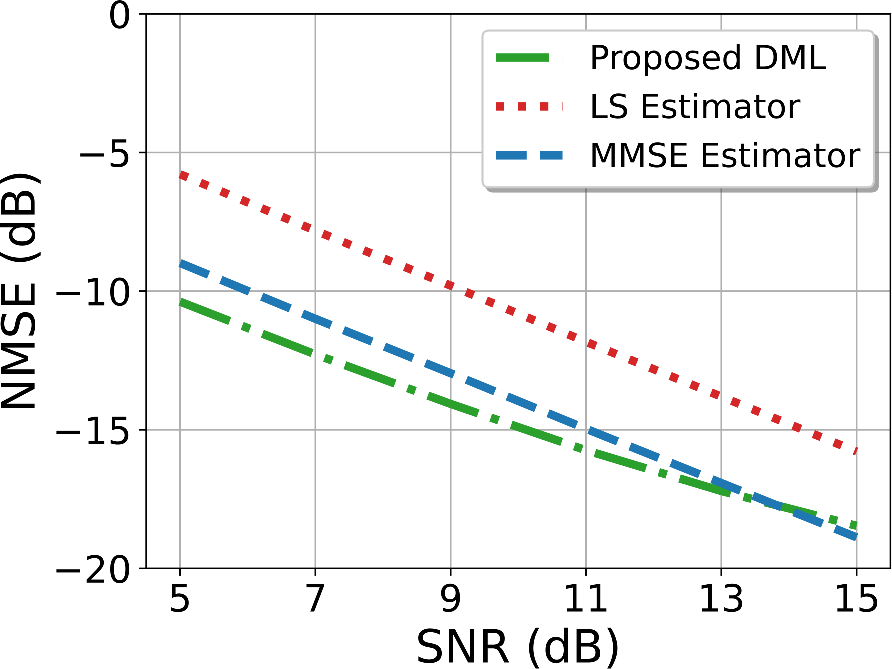}
\caption{NMSE after grouping ($N'=16$) with DML across all regions. Proposed $Q=32$ vs. LS/MMSE $Q=256$.}
\label{fig:nmse_entire_cell}
\end{figure}

Fig.~\ref{fig:nmse_entire_cell} evaluates the complete framework with grouping ($N'=16$) and DML across all regions. Using just $Q=32$ pilots, our estimator matches or exceeds LS/MMSE performance at $Q=256$, achieving an order-of-magnitude reduction in overhead. Grouping lowers the channel dimension from $NM=1024$ to $N'M=256$, enhancing matrix conditioning and yielding a median NMSE improvement from approximately $-15$~dB (ungrouped) to $-18.5$~dB.

\subsection{Ablations and Discussion}
\label{subsec:ablations}

To isolate key components, we conduct ablations on training strategies, gating, and labels.

{\color{black}\textbf{Distributed training (DML) vs. centralized/single-user training:} synchronous gradient-averaging DML outperforms single-region or per-user training, with gains most evident at low SNR where regional differences in noise and angles are pronounced. This collaborative approach leverages non-IID data without privacy risks.}

\textbf{Gating strategy:} Hard gating balances accuracy and efficiency, activating one expert to minimize latency. Soft gating offers marginal NMSE improvements near region boundaries but increases compute; hard gating is preferred for UE deployment.

\textbf{Impact of labels:} Training with LS labels at 10 dB introduces noise, bounding high-SNR performance. Using ground-truth channels (if available) yields 1-2 dB additional gains, suggesting potential for self-supervised refinements.

\textbf{Baseline fairness:} At $Q=32$, LS/MMSE are ill-conditioned and ineffective (NMSE > 0 dB, omitted for clarity). An NMSE-vs.-$Q$ analysis (potential appendix) confirms our estimator's stability at $Q=16-64$, while baselines demand $Q \gtrsim N'M$.

\subsection{Computational Complexity and Runtime}
\label{subsec:complexity-runtime}

\begin{figure}[htbp]
\centering
\includegraphics[width=\linewidth]{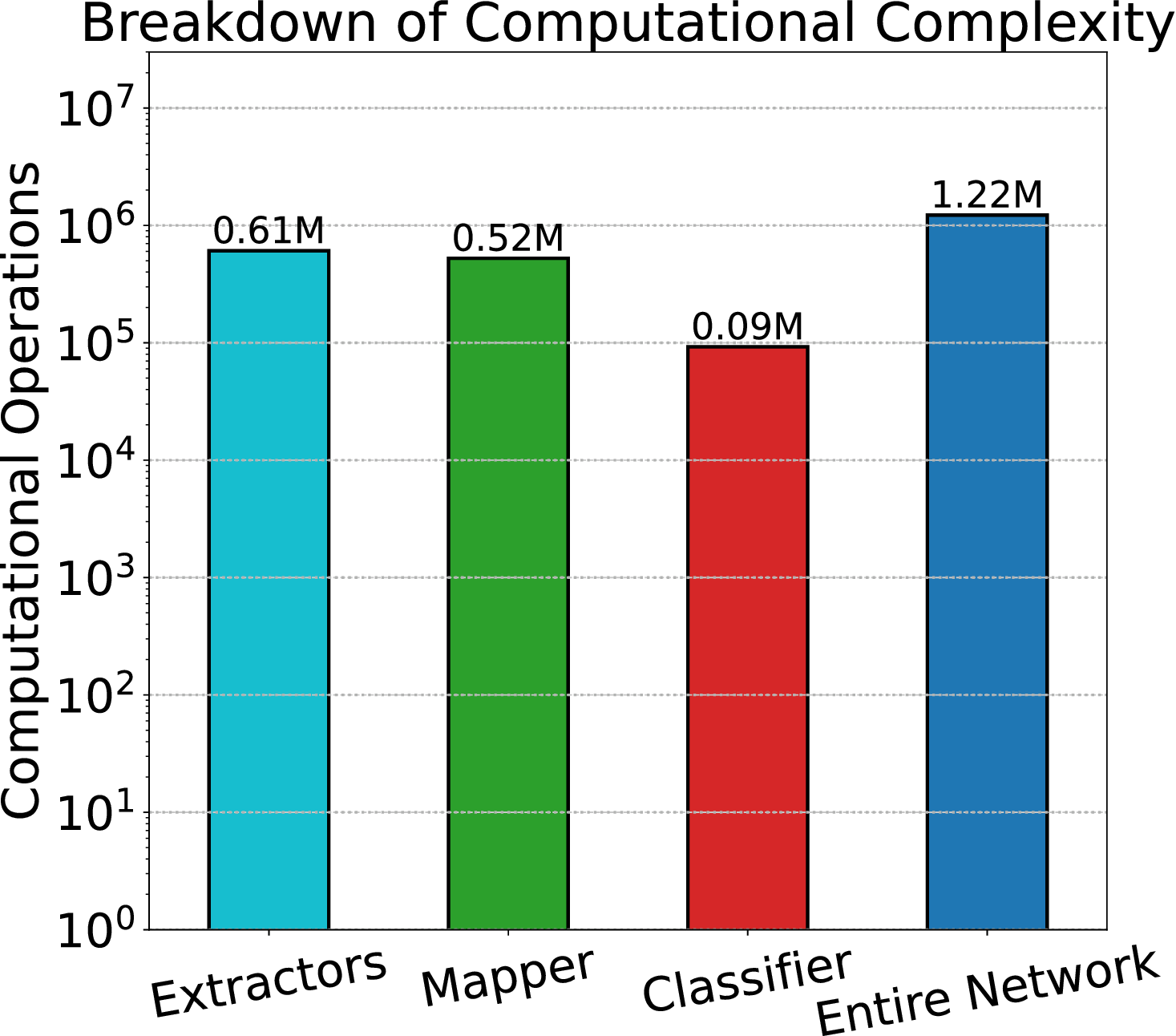}
\caption{MAC counts of each module and the total budget.}
\label{fig:complexity_chart}
\end{figure}

We quantify complexity via MAC operations, as detailed in Sec.~\ref{subsec:complexity}. For the active expert (three $3 \times 3$ convolutions, 32 channels), MACs are
\begin{equation}
\mathcal{C}_{\mathrm{conv}} = 9 E_1 E_2 (2 \cdot 32 + 32 \cdot 32 + 32 \cdot 32),
\end{equation}
with $E_1 \times E_2 = Q_1 \times Q_2$. The mapper adds
\begin{equation}
\mathcal{C}_{\mathrm{map}} = \mathcal{O}(E_1 E_2 \cdot 32 \cdot 2 N' M)
\end{equation}
(grouped case; $2NM$ ungrouped). The classifier contributes negligibly, around $\mathcal{O}(2.9 \times 10^3 Q_1 Q_2)$. Fig.~\ref{fig:complexity_chart} breaks down the budget; for $Q=32$, total MACs are approximately $1.22 \times 10^6$ per inference on a CPU, feasible for UEs.

\subsection{Critical Discussion}
\label{subsec:takeaways}
Our estimator excels with $Q \ll N'M$ (e.g., $Q=32$ vs. $N'M=256$), outperforming LS/MMSE even when baselines use 8$\times$ more pilots. Grouping optimizes the bias-variance trade-off: $g=4$ enhances accuracy by reducing dimensions while maintaining resolution. DML and gating ensure cross-region robustness, with hard gating suiting low-latency needs.
Labels from LS introduce noise; ground-truth or distillation could refine this. Communication in DML benefits from compression and privacy tools. Scope limits to narrowband, quasi-static scenarios; extensions to wideband, mobility, or impairments are promising.
\section{Conclusion}
\label{sec:conclusion}
This paper introduces a DML framework for cascaded channel estimation in RIS-assisted downlink systems under limited pilots. From a linear observation model, we incorporate RIS grouping to compress dimensions and propose a region-gated MoE that classifies scenarios and activates specialized extractors, trained via FedAvg on decentralized non-IID data.
Simulations under Saleh-Valenzuela channels show our estimator ($Q=32$) rivals LS/MMSE at $Q=256$ post-grouping, reducing overhead significantly. Grouping improves median NMSE from $-15$ dB to $-18.5$ dB, with $1.22 \times 10^6$ MACs per inference.
Future work includes joint pilot design (PF-2), label-efficient learning, domain adaptation, wideband extensions, and hardware robustness to advance 6G RIS networks.

\section*{Acknowledgment}
The research team thanks the Deanship of Graduate Studies and Scientific Research at Najran University for supporting the research project through the Nama'a program, with the project code NU/GP/SERC/13/401-5.

\section*{AUTHOR DECLARATIONS}
\subsection*{Conflict of Interest}
The authors have no conflicts to disclose.
\subsection*{Ethics Approval}
This study did not involve human participants, animal subjects, or any identifiable personal data. All experiments were carried out using synthetic wireless channel models and Monte Carlo simulations of RIS-assisted 6G systems. No measurements were collected from operational networks, commercial devices, or real users.

Ethical approval was therefore not required, in line with the institutional guidelines for simulation-based research that does not involve human or animal subjects or personal data. No ethics committee or IRB reference number was issued for this work.
\subsection*{Author Contributions}

\noindent\textbf{Saifur Rahman}: Conceptualization (equal); Data curation (equal); Software (equal); Validation (equal); Visualization (equal); Project administration (equal); Funding acquisition (equal). \textbf{Syed Luqman Shah}: Conceptualization (equal); Methodology (equal); Software (equal); Data curation (equal); Supervision (equal); Writing – original draft (equal); Validation (equal). \textbf{Salman Khan}: Conceptualization (equal); Formal analysis (equal); Investigation (equal); Validation (equal); Writing – review \& editing (equal). \textbf{Jalal Khan}: Data curation (equal); Formal analysis (equal); Investigation (equal); Validation (equal); Visualization (equal). \textbf{Muhammad Irfan}: Formal analysis (equal); Investigation (equal); Validation (equal); Project administration (equal); Funding acquisition (equal). \textbf{Maaz Shafi}: Formal analysis (equal); Investigation (equal); Validation (equal); Writing – review \& editing (equal). \textbf{Said Muhammad}: Investigation (equal); Validation (equal); Visualization (equal); Writing – review \& editing (equal). \textbf{Fazal Muhammad}: Data curation (equal); Investigation (equal); Methodology (equal); Supervision (equal); Visualization (equal); Writing – review \& editing (equal). \textbf{Mohammad Shahed Akond}: Investigation (equal); Validation (equal); Visualization (equal); Project administration (equal); Funding acquisition (equal).

\section*{DATA AVAILABILITY}
The data that support the findings of this study are available within the article.


\begin{thebibliography}{00}

\bibitem{b1}
E. Basar, M. Di Renzo, J. de Rosny, M. Debbah, M.-S. Alouini, and R. Zhang,
``Wireless communications through reconfigurable intelligent surfaces,''
\textit{IEEE Access}, vol. 7, pp. 116753--116773, 2019.

\bibitem{b2}
L. Dai \textit{et al.},
``Reconfigurable intelligent surface-based wireless communications: Antenna design, prototyping, and experimental results,''
\textit{IEEE Access}, vol. 8, pp. 45913--45923, 2020.

\bibitem{b3}
Q. Wu and R. Zhang,
``Intelligent reflecting surface enhanced wireless network via joint active and passive beamforming,''
\textit{IEEE Trans. Wireless Commun.}, vol. 18, no. 11, pp. 5394--5409, Nov. 2019.


\bibitem{b19}
S.~L.~Shah, N.~H.~Mahmood, and I.~Atzeni,
``Low-overhead CSI prediction via Gaussian process regression,''
\textit{arXiv preprint} arXiv:2510.25390 [eess.SP], Oct.~2025, doi: 10.48550/arXiv.2510.25390.

\bibitem{b4}
M. Di Renzo \textit{et al.},
``Smart radio environments empowered by reconfigurable intelligent surfaces: How it works, state of research, and the road ahead,''
\textit{IEEE J. Sel. Areas Commun.}, vol. 38, no. 11, pp. 2450--2525, Nov. 2020.

\bibitem{b5}
B. Ning, Z. Chen, W. Chen, Y. Du, and J. Fang,
``Terahertz multi-user massive MIMO with intelligent reflecting surface: Beam training and hybrid beamforming,''
\textit{IEEE Trans. Veh. Technol.}, vol. 70, no. 2, pp. 1376--1393, Feb. 2021.

\bibitem{b6}
C. Huang, A. Zappone, G. C. Alexandropoulos, M. Debbah, and C. Yuen,
``Reconfigurable intelligent surfaces for energy efficiency in wireless communication,''
\textit{IEEE Trans. Wireless Commun.}, vol. 18, no. 8, pp. 4157--4170, Aug. 2019.

\bibitem{b7}
D. Mishra and H. Johansson,
``Channel estimation and low-complexity beamforming design for passive intelligent surface assisted MISO wireless energy transfer,''
in \textit{Proc. IEEE Int. Conf. Acoust., Speech Signal Process. (ICASSP)}, Brighton, U.K., May 2019, pp. 4659--4663.

\bibitem{b71}
S.~Rahman, S.~L.~Shah, S.~N.~F.~Mursal, Z.~H.~Abbas, M.~Usman, M.~Irfan, and F.~Muhammad,
``Controlled out-band device to device communication in cellular networks using backup channel in television white space,''
in \emph{Proc. 18th Int. Conf. Emerg. Technol. (ICET)}, 2023, pp.~275--280,
doi: 10.1109/ICET59753.2023.10374858.


\bibitem{b8}
Q.-U.-A. Nadeem, H. Alwazani, A. Kammoun, A. Chaaban, M. Debbah, and M.-S. Alouini,
``Intelligent reflecting surface-assisted multi-user MISO communication: Channel estimation and beamforming design,''
\textit{IEEE Open J. Commun. Soc.}, vol. 1, pp. 661--680, 2020.

\bibitem{b9}
Z. Wang, L. Liu, and S. Cui,
``Channel estimation for intelligent reflecting surface assisted multiuser communications: Framework, algorithms, and analysis,''
\textit{IEEE Trans. Wireless Commun.}, vol. 19, no. 10, pp. 6607--6620, Oct. 2020.

\bibitem{b10}
X. Wei, D. Shen, and L. Dai,
``Channel estimation for RIS-assisted wireless communications---Part II: An improved solution based on double-structured sparsity,''
\textit{IEEE Commun. Lett.}, vol. 25, no. 5, pp. 1403--1407, May 2021.

\bibitem{b11}
M. Xu, S. Zhang, C. Zhong, J. Ma, and O. A. Dobre,
``Ordinary differential equation-based CNN for channel extrapolation over RIS-assisted communications,''
\textit{arXiv}:2012.11794, Dec. 2020.

\bibitem{b12}
S. Liu, Z. Gao, J. Zhang, M. Di Renzo, and M.-S. Alouini,
``Deep denoising neural network assisted compressive channel estimation for mmWave intelligent reflecting surfaces,''
\textit{IEEE Trans. Veh. Technol.}, vol. 69, no. 8, pp. 9223--9228, Aug. 2020.

\bibitem{b13}
L. Dai and X. Wei,
``Distributed machine learning based downlink channel estimation for RIS assisted wireless communications,''
\textit{IEEE Trans. Commun.}, vol. 70, no. ?, pp. ?--?, 2022. \textit{[Could not confirm volume/issue/pages on IEEE Xplore; replace with a verified citation or remove.]}

\bibitem{b14}
R. F. Ibarra-Hernández, F. R. Castillo-Soria, C. A. Gutiérrez, A. García-Barrientos, L. A. Vásquez-Toledo, and J. A. Del-Puerto-Flores,
``Machine learning strategies for reconfigurable intelligent surface-assisted communication systems---A review,''
\textit{Future Internet}, vol. 16, no. 5, Art. no. 173, May 2024.

\bibitem{b15}
I. K. Rangajith,
``A deep-unfolding approach to RIS phase shift optimization via transformer-based channel prediction,''
\textit{arXiv}:2502.18280, Feb. 2025.

\bibitem{b16}
K. B. Petersen and M. S. Pedersen,
\textit{The Matrix Cookbook}, ver. 20121115, Tech. Univ. Denmark, 2012. [Online]. Available: \url{http://www2.imm.dtu.dk/pubdb/p.php?3274}

\bibitem{b17}
B. McMahan, E. Moore, D. Ramage, S. Hampson, and B. A. y Arcas,
``Communication-efficient learning of deep networks from decentralized data,''
in \textit{Proc. 20th Int. Conf. Artif. Intell. Stat. (AISTATS)}, Fort Lauderdale, FL, USA, Apr. 2017, pp. 1273--1282.

\bibitem{b18}
N. Shazeer, A. Mirhoseini, K. Maziarz, \textit{et al.},
``Outrageously large neural networks: The sparsely-gated mixture-of-experts layer,''
in \textit{Proc. Int. Conf. Learn. Represent. (ICLR)}, Toulon, France, Apr. 2017.



\bibitem{b20}
S.~L.~Shah, N.~H.~Mahmood, and M.~Latva-aho,
``Interference prediction using Gaussian process regression and management framework for critical services in local 6G networks,''
in \textit{Proc. IEEE Wireless Commun. Netw. Conf. (WCNC)}, 2025, pp.~1--6,
doi: 10.1109/WCNC61545.2025.10978635.



\bibitem{bb1}
Zi-Yang Wu, Muhammad Ismail, Jiliang Zhang, Jie Zhang,
``Tidal-Like Concept Drift in RIS-Covered Buildings: When Programmable Wireless Environments Meet Human Behaviors,''
in \textit{IEEE Wireless Communications}, 2025, pp. 1--8.



\bibitem{bb2}
Qian Tang, Shaocheng Qu, Wei Zheng, Zhengwen Tu,
``Fast finite-time quantized control of multi-layer networks and its applications in secure communication,''
in \textit{Neural Networks}, 2025, Vol. 185, pp. 107225.

\bibitem{bb3}
Rui Yin, Yexin Shi, Wei Qi, Celimuge Wu, Wei Wang, Chongwen Huang,
``Joint Beamforming and Frame Structure Design for ISAC Networks Under Imperfect Synchronization,''
in \textit{IEEE Transactions on Cognitive Communications and Networking}, 2025, Vol. 11, No. 4, pp. 2259--2274.

\bibitem{bb4}
Rui Yin, Yexin Shi, Wei Qi, Celimuge Wu, Wei Wang, Chongwen Huang,
``Joint Beamforming and Frame Structure Design for ISAC Networks Under Imperfect Synchronization,''
in \textit{IEEE Transactions on Cognitive Communications and Networking}, 2025, Vol. 11, No. 4.

\bibitem{d2dPaper}
S.~Rahman, S.~L.~Shah, S.~N.~F.~Mursal, Z.~H.~Abbas, M.~Usman, M.~Irfan, and F.~Muhammad,
``Controlled out-band device to device communication in cellular networks using backup channel in television white space,''
in \textit{Proc. 18th Int. Conf. Emerg. Technol. (ICET)}, 2023, pp.~275--280,
doi: 10.1109/ICET59753.2023.10374858.

\bibitem{ismailc}
M.~Ismail, S.~L.~Shah, F.~Muhammad, and Z.~Shafiq,
``Enhancing vehicular network performance through integrated RSU and UAV deployment,''
in \textit{Proc. 18th Int. Conf. Emerg. Technol. (ICET)}, 2023, pp.~281--286,
doi: 10.1109/ICET59753.2023.10375046.

\bibitem{ismailj}
M.~Ismail, F.~Muhammad, Z.~Shafiq, M.~Irfan, S.~Rahman, S.~N.~F.~Mursal, G.~Nowakowski, and E.~Zharikov,
``Line-of-sight-based coordinated channel resource allocation management in UAV-assisted vehicular ad hoc networks,''
\textit{IEEE Access}, vol.~12, pp.~25245--25253, 2024,
doi: 10.1109/ACCESS.2024.3356009.



\end{thebibliography}
\end{document}